\documentclass[a4paper,showpacs,showkeys,prl]{revtex4}

\usepackage{amsmath}
\usepackage{amsfonts}
\usepackage[all]{xy}

\newcommand{\bk}{\emph{\textbf{k}}}
\newcommand{\bq}{\emph{\textbf{q}}}
\newcommand{\br}{\emph{\textbf{r}}}

\newcommand{\rd}{\mathrm{d}}
\newcommand{\qpx}{q'_x\phantom{}}
\newcommand{\qpy}{q'_y\phantom{}}
\newcommand{\unmedio}{{\textstyle\frac{1}{2}}}
\newcommand{\tresmedio}{{\textstyle\frac{3}{2}}}

\begin{document}

\title{Exact solution for two-dimensional Coulomb matrix elements}

\author{Jaime \surname{Zaratiegui Garc{\'\i}a}}
 \affiliation{Department of Physical Sciences/Theoretical Physics,
  P.O. Box 3000, FIN-90014, University of Oulu, Finland} %
 \email{jaime.zaratiegui@oulu.fi}

\begin{abstract}
Exact analytic expression is derived for the matrix elements of the
Coulomb interaction in two dimensions in the form of a closed finite
sum expression. The orthonormal complete set of eigenfunctions of the
harmonic oscillator $\left.\right|n_xn_y\left>\right.$ is used as the
basis for spanning real space. Several recurrence relations have been
found in order to simplify the task of calculating the usually vast
amount of elements required for any computer simulation.
\end{abstract}

\maketitle

The choice of the anisotropic orthonormal complete set of
eigenfunctions of the harmonic oscillator is due to the ubiquity and
special symmetric properties of this particular problem in the field
of quantum mechanics. The projection of this basis into real space is
\begin{eqnarray}
\left<xy\left|\right.n_xn_y\right> = \psi_{n_xn_y}(x,y) & = &
\left(\frac{1}{a_xa_y}\right)^{1/2} \frac{1}{\sqrt{\pi}}
\left(\frac{1}{2^{n_x}n_x!} \frac{1}{2^{n_y}n_y!}\right)^{1/2}
\nonumber \\
 &   & e^{-\frac{1}{2}x^2/a_x^2}H_{n_x}(x/a_x)
 e^{-\frac{1}{2}y^2/a_y^2}H_{n_y}(y/a_y), \\
\left<n'_xn'_y\left|\right.n_xn_y\right> & = &
 \delta_{n_xn'_x}\delta_{n_yn'_y}.
\end{eqnarray}
Without any loss of generality we can assume that
\begin{equation}
a_x>a_y.
\end{equation}
Just the case $a_x = a_y$ will fall apart from our supposition. There
is no problem in treating this case apart, as it is even simpler to
calculate and can be derived from our results. The eccentricity
parameter $\gamma$ is defined as 
\begin{eqnarray}
&\gamma = \frac{a_y}{a_x},& \\
& 0 < \gamma < 1.&
\end{eqnarray}

It is of great interest to express the wavefunctions and Coulomb
potential as their two-dimensional Fourier transforms
\begin{eqnarray}
\psi_\lambda(\br) & = & \frac{1}{2\pi}\int \phi_\lambda(\bq)
e^{-i\bq\cdot\br}\,\mathrm{d}\bq, \\
V(\br) & = & \frac{1}{2\pi}\int \tilde{V}(\bq) e^{-i\bq\cdot\br}
\,\mathrm{d}\bq.
\end{eqnarray}
Under this notation, the matrix elements take a different expression
\begin{equation}
\mathcal{V}_{\lambda_3\lambda_4}^{\lambda_1\lambda_2} =
\frac{1}{2\pi} \int
\phi^\ast_{\lambda_1}(\bq_1)\phi_{\lambda_4}(\bq_1-\bq)
\phi^\ast_{\lambda_2}(\bq_2)\phi_{\lambda_3}(\bq_2+\bq)
\tilde{V}(\bq)\,\mathrm{d}\bq_1\mathrm{d}\bq_2\mathrm{d}\bq,
\end{equation}
where $\lambda_i$ represents the set of quantum numbers
$\left\{n^i_x,n^i_y\right\}$ for the $i$-th particle.
\begin{equation}
\mathcal{V}_{\lambda_3\lambda_4}^{\lambda_1\lambda_2} =
\frac{1}{2\pi} \int C^{\lambda_1}_{\lambda_4}(\bq)
D^{\lambda_2}_{\lambda_3}(\bq) \tilde{V}(\bq)\,\mathrm{d}\bq,
\end{equation}
where
\begin{eqnarray}
C^\lambda_{\lambda^\prime}(\bq) & = &
\int\phi_\lambda^\ast(\bk)\phi_{\lambda^\prime}(\bk-\bq)\,\mathrm{d}\bk
= \int \psi_\lambda^\ast(\br) \psi_{\lambda^\prime}(\br)
e^{-i\bq\cdot\br} \,\mathrm{d} \br, \\
D^\lambda_{\lambda^\prime}(\bq) & = &
\int \phi_\lambda^\ast(\bk) \phi_{\lambda^\prime}(\bk+\bq)
\,\mathrm{d}\bk = \int \psi_\lambda^\ast(\br)
\psi_{\lambda^\prime}(\br) e^{i\bq\cdot\br}\,\mathrm{d} \br \nonumber \\
 & = & C^\lambda_{\lambda^\prime}(-\bq).
\end{eqnarray}

The calculation of $C_{\lambda\lambda^\prime}(\bq)$ is quite
straightforward. It can be separated in both two dimensions, as they
are (almost) symmetrical. The calculation proceeds as follows
(normalization constants omitted for clarity)
\begin{eqnarray}
\int_{-\infty}^{\infty} e^{-x^2/a_x^2} H_{n_x^1}(x/a_x)
H_{n_x^4}(x/a_x) e^{-i q_x x} \, \mathrm{d}x & = & a_x
 \int_{-\infty}^{\infty} e^{-x'^2} H_{n_x^1}(x') H_{n_x^4}(x') e^{-i
 q_x a_x x'} \, \mathrm{d}x' \nonumber \\
 & = & a_x \int_{-\infty}^{\infty}  e^{-x'^2} H_{n_x^1}(x')
 H_{n_x^4}(x') e^{-i\qpx  x'} \, \mathrm{d}x',
\end{eqnarray}
where $x'= a_x x$ and $\qpx  = a_x q_x$. This equation can be seen as the
Fourier transform of the product of two functions:
\begin{equation}
\mathcal{F}\left[ f_{n_x^1} f_{n_x^4} \right],
\end{equation}
being
\begin{equation}
 f_{n} (x) = e^{-x^2/2} H_{n} (x). 
\end{equation}
Making use of the properties of the Fourier transform, in particular
the one that states that the transform of the product of two functions
is the convolution of their transforms, i.e.
\begin{equation}
\mathcal{F}[fg] = \mathcal{F}[f] \ast \mathcal{F}[g].
\end{equation} 
In our particular case, the Fourier transform of the function $f_n(x)$ has a
very simple form as it is an eigenfunction of this operator with
associated eigenvalue $i^n$:
\begin{equation}
\mathcal{F}[f_n(x)] (q_x) = i^n f_n(q_x).
\end{equation} 
It only remains now to calculate the convolution product, defined as
\begin{equation}
f \ast g \equiv \int_{-\infty}^{\infty} f(\tau) g(t-\tau) \,
\mathrm{d} \tau = \int_{-\infty}^{\infty} g(\tau) f(t-\tau) \,
\mathrm{d} \tau.
\end{equation}
In our case, we have to calculate the following integral
\begin{equation}
\mathcal{F}[f_{n_x^1} f_{n_x^4}] = i^{n_x^1+n_x^4} \int_{-\infty}^{\infty} e^{-x'^2/2} H_{n_x^1} (x')
e^{-(\qpx -x')^2/2} H_{n_x^4} (\qpx -x') \, \mathrm{d} x'.
\end{equation}
Let us first do the following change of variables:
\begin{equation}
\left\{ \begin{array}{ll}
x' = u + \qpx /2 & \qpx  = 2x' -  2 u \\
\mathrm{d}x' = \mathrm{d}u & (-\infty, \infty) \rightarrow  (-\infty, \infty)
\end{array} \right. .
\end{equation}
Thus, choosing $n_{x+}^{14} = \max (n_x^1, n_x^4)$ and $n_{x-}^{14} =
\min (n_x^1, n_x^4)$ , the last integral is transformed to
\begin{eqnarray}
 & & \int_{-\infty}^{\infty} e^{-(u+\qpx /2)^2/2} H_{n_{x-}^{14}} (u+\qpx /2)
e^{-(\qpx -u-\qpx /2)^2/2} H_{n_{x+}^{14}} (\qpx -u-\qpx /2) \,
\mathrm{d} u \nonumber \\
 & = & e^{-\qpx ^2/4} \int_{-\infty}^{\infty} e^{-u^2} H_{n_{x-}^{14}}
 (u+\qpx /2) (-1)^{n_{x+}^{14}}H_{n_{x+}^{14}} (u-\qpx /2) \,
 \mathrm{d} u, \label{eq:gradsh-1}
\end{eqnarray}
which according to Gradshteyn \& Ryzhik Eq. 7.377 in page 797
\begin{equation}
\int_{-\infty}^{\infty} e^{-x^2} H_m (x+y) H_n (x+z) \, dx = 2^n
\pi^{1/2} m! z^{n-m} L_{m}^{n-m} (-2yz) \, \, [m \le n],
\end{equation}
the result of integral~(\ref{eq:gradsh-1}) is
\begin{equation}
 2^{n_{x+}^{14}} \pi^{1/2} n_{x-}^{14}! \left(  -\frac{a_xq_x}{2}
 \right)^{n_{x+}^{14}-n_{x-}^{14}}
 L_{n_{x-}^{14}}^{n_{x+}^{14}-n_{x-}^{14}} (a_x^2q_x^2/2).
\end{equation}
Collecting all terms together, we can write the final result for
$C(\bq)$:
\begin{eqnarray}
C_{n_x^4 n_y^4}^{n_x^1 n_y^1}(q_x, q_y) & = &  \left(
\frac{1}{a_xa_y \pi} \right) \left( \frac{1}{2^{n_x^1}n_x^1!}
\frac{1}{2^{n_y^1}n_y^1!} \frac{1}{2^{n_x^4}n_x^4!}
\frac{1}{2^{n_y^4}n_y^4!} \right)^{1/2} \nonumber \\
 &   & i^{n_x^1+n_y^1+n_x^4+n_y^4} \nonumber \\
 &  & a_xa_y 2^{n_{x+}^{14}} 2^{n_{y+}^{14}}
 \pi n_{x-}^{14}! n_{y-}^{14}!
 (-1)^{n_{x+}^{14} + {n_{y+}^{14}} } \nonumber \\
 &   & e^{-a_x^2 q_x^2/4} \left(  -\frac{a_xq_x}{2}
 \right)^{n_{x+}^{14}-n_{x-}^{14}}
 L_{n_{x-}^{14}}^{n_{x+}^{14}-n_{x-}^{14}} (a_x^2q_x^2/2) \nonumber \\ 
 &   &  e^{-a_y^2 q_y^2/4} \left(  -\frac{a_yq_y}{2}
 \right)^{n_{y+}^{14}-n_{y-}^{14}}
 L_{n_{y-}^{14}}^{n_{y+}^{14}-n_{y-}^{14}} (a_y^2q_y^2/2) \nonumber \\
 & = & (-1)^{n_{x-}^{14} + {n_{y-}^{14}}} \left( \frac{2^{n_{x+}^{14}}}{n_{x+}^{14}!}
\frac{n_{x-}^{14}!}{2^{n_{x-}^{14}}} \frac{2^{n_{y+}^{14}}}{n_{y+}^{14}!}
\frac{n_{y-}^{14}!}{2^{n_{y-}^{14}}} \right)^{1/2} \nonumber \\
 &   & i^{n_x^1+n_y^1+n_x^4+n_y^4} \nonumber \\
 &   & e^{-\frac{1}{4}(a_x^2q_x^2+a_y^2q_y^2)} \left(
 \frac{a_xq_x}{2} \right)^{n_{x+}^{14}-n_{x-}^{14}}  \left(
 \frac{a_yq_y}{2} \right)^{n_{y+}^{14}-n_{y-}^{14}} \nonumber  \\
 &   & L_{n_{x-}^{14}}^{n_{x+}^{14}-n_{x-}^{14}} (a_x^2q_x^2/2)
 L_{n_{y-}^{14}}^{n_{y+}^{14}-n_{y-}^{14}} (a_y^2q_y^2/2).
\end{eqnarray}
In the same fashion we calculate the function $D$:
\begin{eqnarray}
D^{n_x^2 n_y^2}_{n_x^3 n_y^3}(q_x, q_y) & = &
(-1)^{n_{x+}^{23} + n_{y+}^{23}} \left(
\frac{2^{n_{x+}^{23}}}{n_{x+}^{23}!}
\frac{n_{x-}^{23}!}{2^{n_{x-}^{23}}} \frac{2^{n_{y+}^{23}}}{n_{y+}^{23}!}
\frac{n_{y-}^{23}!}{2^{n_{y-}^{23}}} \right)^{1/2} \nonumber \\
 &   & i^{n_x^2+n_y^2+n_x^3+n_y^3} \nonumber \\
 &   & e^{-\frac{1}{4}(a_x^2q_x^2+a_y^2q_y^2)} \left(
 \frac{a_xq_x}{2} \right)^{n_{x+}^{23}-n_{x-}^{23}}  \left(
 \frac{a_yq_y}{2} \right)^{n_{y+}^{23}-n_{y-}^{23}} \nonumber \\
 &   & L_{n_{x-}^{23}}^{n_{x+}^{23}-n_{x-}^{23}} (a_x^2q_x^2/2)
 L_{n_{y-}^{23}}^{n_{y+}^{23}-n_{y-}^{23}} (a_y^2q_y^2/2).
\end{eqnarray}

The Fourier transform of the Coulombic potential is
\begin{eqnarray}
\tilde{V}(q) & = & \int_0^\infty V(r) J_0(rq) r\,\mathrm{d}r=
\xi \frac{1}{q} = \xi \frac{2}{\sqrt{\pi}} \int_0^\infty
e^{-(q_x^2+q_y^2)u^2}\,\rd u\\
 & = & \xi \frac{2}{\sqrt{\pi}} \int_0^\infty  e^{-(a_x^2q_x^2 +
 a_x^2 a_y^2 q_y^2/a_y^2) u^2/a_x^2}\,\rd u \\
 & = & \xi a_x \frac{2}{\sqrt{\pi}} \int_0^\infty e^{-(\qpx^2 +
 \qpy^2/\gamma^2)u^2} \,\rd u \\
V(r) & = & \frac{\xi}{r}.
\end{eqnarray}
Where we have substituted the dimensionless momentum variables,
\begin{eqnarray}
q_x^{\prime} & = & a_x q_x, \\
q_y' & = & a_y q_y.
\end{eqnarray}

The integration can be now performed. Making use of the expansion of
the Laguerre polynomials as
\begin{equation}
L_n^l(x)=\sum_{k=0}^{n}\frac{(-1)^k}{k!}\binom{n+l}{n-k} x^k,
\end{equation}
let us focus on the $q_x$ and $q_y$ dependent part. Let us also forget
about any multiplicative constants and summation
indices, we can write the matrix element as:
\begin{eqnarray}
\mathcal{V}^{n_x^1n_y^1n_x^2n_y^2}_{n_x^3n_y^3n_x^4n_y^4} & = &
\frac{1}{2\pi}
(-1)^{n_{x-}^{14} + n_{y-}^{14} + n_{x+}^{23} + n_{y+}^{23}} \left(
\frac{2^{n_{x+}^{14}}}{n_{x+}^{14}!}
\frac{n_{x-}^{14}!}{2^{n_{x-}^{14}}}
\frac{2^{n_{y+}^{14}}}{n_{y+}^{14}!} 
\frac{n_{y-}^{14}!}{2^{n_{y-}^{14}}}
\frac{2^{n_{x+}^{23}}}{n_{x+}^{23}!}
\frac{n_{x-}^{23}!}{2^{n_{x-}^{23}}}
\frac{2^{n_{y+}^{23}}}{n_{y+}^{23}!} 
\frac{n_{y-}^{23}!}{2^{n_{y-}^{23}}} \right)^{1/2} \nonumber \\
 &   & (-1)^{s_x + s_y + n_{x-}^{14} + n_{y-}^{14} + n_{x-}^{23} +
 n_{y-}^{23} } \nonumber \\
 &   & \xi a_x \frac{2}{\sqrt{\pi}} \nonumber \\
 &   & \sum_{k_x = 0}^{n_{x-}^{14}} \frac{(-1)^{k_x}}{k_x!}
 \binom{n_{x+}^{14}}{n_{x-}^{14}-k_x}
 \sum_{k'_x = 0}^{n_{x-}^{23}} \frac{(-1)^{k'_x}}{k'_x!}
 \binom{n_{x+}^{14}}{n_{x-}^{14}-k'_x} \nonumber \\
 &   & \sum_{k_y = 0}^{n_{y-}^{14}} \frac{(-1)^{k_y}}{k_y!}
 \binom{n_{y+}^{14}}{n_{y-}^{14}-k_y}
 \sum_{k'_y = 0}^{n_{y-}^{23}} \frac{(-1)^{k'_y}}{k'_y!}
 \binom{n_{y+}^{14}}{n_{y-}^{14}-k'_y} \nonumber \\
 &   & \left( \frac{1}{2} \right)^{s_x + s_y + k_x + k'_x + k_y + k'_y}
 \int_0^\infty\rd u \int_{-\infty}^{\infty} \frac{\rd q'_x}{a_x}
 e^{-(u^2+1/2)\qpx^2}  \qpx^{(2s_x+2k_x+2k'_x)} \nonumber \\
 &   & \int_{-\infty}^{\infty} \frac{\rd q'_y}{a_y}
 e^{-(u^2/\gamma^2+1/2)\qpy^2} \qpy^{(2s_y+2k_y+2k'_y)} \label{eq:int}
\end{eqnarray}
The last three integrals can be solved quite easily. First, the ones
depending in $\qpx$ and $\qpy$ lead to:
\begin{eqnarray}
\int_{-\infty}^{\infty} e^{-(u^2+1/2)\qpx^2}
\qpx^{(2s_x+2k_x+2k'_x)}\,\rd q'_x & = & 2^{a+1/2} \Gamma \left(
a + 1/2 \right) \frac{1}{(1+2u^2)^{a+1/2}} \\
\int_{-\infty}^{\infty}
 e^{-(u^2/\gamma^2+1/2)\qpy^2} \qpy^{(2s_y+2k_y+2k'_y)}\,\rd q'_y & =
 & 2^{b+1/2} \Gamma \left(b + 1/2 \right)
 \frac{1}{(1+2u^2/\gamma^2)^{b+1/2}}
\end{eqnarray}
where $a = s_x+k_x+k'_x$ and $b = s_y+k_y+k'_y$. We have
already included the condition for the integral to be
non-zero:
\begin{eqnarray}
2s_x & = & |n_x^1-n_x^4|+|n_x^2-n_x^3| =
n_{x+}^{14} - n_{x-}^{14} + n_{x+}^{23} - n_{x-}^{23} ,\\
2s_y & = & |n_y^1-n_y^4|+|n_y^2-n_y^3| =
n_{y+}^{14} - n_{y-}^{14} + n_{y+}^{23} - n_{y-}^{23} , \\
s_x, s_y & \in & \mathbb{N}.
\end{eqnarray}
Now, the $u$-integration can be performed.
\begin{eqnarray}
\int_0^{\infty}\frac{1}{(1+2u^2/\gamma^2)^{b+1/2}}
\frac{1}{(1+2u^2)^{a+1/2}} \, \rd u & = & \frac{1}{2}
\left( \frac{2}{\gamma^2} \right)^{-1/2} \mathrm{B}\left( \frac{1}{2},
a + b + \frac{1}{2} \right) \nonumber \\ 
&   & \times \phantom{}_2F_1 \left( a + \frac{1}{2}, \frac{1}{2}; a +
b + 1, 1 - \gamma^2 \right) \label{eq:iu}
\end{eqnarray}
which was solved with the help of Gradshteyn \& Ryzhik Integral
3.259 3, page 326 (check also the errata)
\begin{equation}
\int_0^\infty x^{\lambda-1} (1 + \alpha x^p)^{-\mu} (1 + \beta x^p)^{-\nu}
\, \rd x = \frac{1}{p} \alpha^{-\lambda/p}
\mathrm{B} \left( \frac{\lambda}{p}, \mu + \nu - \frac{\lambda}{p}
\right) \phantom{} _2F_1 \left( \nu, \frac{\lambda}{p}; \mu + \nu; 1 -
\frac{\beta}{\alpha} \right)
\end{equation}
\begin{displaymath}
\left[ |\arg \alpha| < \pi, \quad |\arg \beta| < \pi, \quad p > 0,
\quad 0 < \mathrm{Re} \, \lambda < 2 \mathrm{Re}(\mu + \nu) \right]
\end{displaymath}
Grouping all terms together, we end up with:
\begin{eqnarray}
\mathcal{V}^{n_x^1n_y^1n_x^2n_y^2}_{n_x^3n_y^3n_x^4n_y^4} & = & \frac{1}{2\pi}
(-1)^{n_{x-}^{14} + n_{y-}^{14} + n_{x+}^{23} + n_{y+}^{23}} \left(
\frac{2^{n_{x+}^{14}}}{n_{x+}^{14}!}
\frac{n_{x-}^{14}!}{2^{n_{x-}^{14}}}
\frac{2^{n_{y+}^{14}}}{n_{y+}^{14}!} 
\frac{n_{y-}^{14}!}{2^{n_{y-}^{14}}}
\frac{2^{n_{x+}^{23}}}{n_{x+}^{23}!}
\frac{n_{x-}^{23}!}{2^{n_{x-}^{23}}}
\frac{2^{n_{y+}^{23}}}{n_{y+}^{23}!} 
\frac{n_{y-}^{23}!}{2^{n_{y-}^{23}}} \right)^{1/2} \nonumber \\
 &   & (-1)^{s_x + s_y + n_{x-}^{14} + n_{y-}^{14} + n_{x-}^{23} +
 n_{y-}^{23} } \nonumber \\
 &   & \left( \frac{1}{2} \right)^{2s_x} \left( \frac{1}{2}
 \right)^{2s_y} \xi a_x \frac{2}{\sqrt{\pi}} \nonumber \\
 &   & \sum_{k_x = 0}^{n_{x-}^{14}} \frac{(-1)^{k_x}}{k_x!}
 \binom{n_{x+}^{14}}{n_{x-}^{14}-k_x}
 \sum_{k'_x = 0}^{n_{x-}^{23}} \frac{(-1)^{k'_x}}{k'_x!}
 \binom{n_{x+}^{14}}{n_{x-}^{14}-k'_x} \nonumber \\
 &   & \sum_{k_y = 0}^{n_{y-}^{14}} \frac{(-1)^{k_y}}{k_y!}
 \binom{n_{y+}^{14}}{n_{y-}^{14}-k_y}
 \sum_{k'_y = 0}^{n_{y-}^{23}} \frac{(-1)^{k'_y}}{k'_y!}
 \binom{n_{y+}^{14}}{n_{y-}^{14}-k'_y} \nonumber \\
 &   & \left( \frac{1}{2} \right)^{k_x + k'_x + k_y + k'_y}
 \frac{1}{a_x} \frac{1}{a_y} 2^{s_x+s_y+k_x+k_y+k'_x+k'_y + 1}
 \Gamma(s_x + k_x + k'_x + 1/2)
 \Gamma(s_y + k_y + k'_y + 1/2) \nonumber \\
 &   & \frac{1}{2} \left( \frac{\gamma^2}{2} \right)^{1/2}
 \mathrm{B}\left( \frac{1}{2}, s_x + s_y + k_x + k_y + k'_x + k'_y +
 \frac{1}{2} \right) \nonumber \\
 &   & \phantom{}_2F_1 \left( s_x + k_x +k'_x + \frac{1}{2},
 \frac{1}{2}; s_x + s_y + k_x + k_y + k'_x + k'_y + 1, 1 -
 \gamma^2 \right).
\end{eqnarray}
After some compactation we get the final result:
\begin{eqnarray}
\mathcal{V}^{n_x^1n_y^1n_x^2n_y^2}_{n_x^3n_y^3n_x^4n_y^4} & = &
\frac{1}{a_x}
\frac{\xi}{\sqrt{2\pi^3}}
(-1)^{s_x + s_y + n_x^2 + n_x^3 + n_y^2 + n_y^3} \left(
\frac{n_{x-}^{14}!}{n_{x+}^{14}!}
\frac{n_{y-}^{14}!}{n_{y+}^{14}!} 
\frac{n_{x-}^{23}!}{n_{x+}^{23}!}
\frac{n_{y-}^{23}!}{n_{y+}^{23}!} 
\right)^{1/2} \nonumber \\
 &   & \sum_{k_x = 0}^{n_{x-}^{14}} \frac{(-1)^{k_x}}{k_x!}
 \binom{n_{x+}^{14}}{n_{x-}^{14}-k_x}
 \sum_{k'_x = 0}^{n_{x-}^{23}} \frac{(-1)^{k'_x}}{k'_x!}
 \binom{n_{x+}^{23}}{n_{x-}^{23}-k'_x} \Gamma(s_x + k_x + k'_x +
 \unmedio)  \nonumber \\
 &   & \sum_{k_y = 0}^{n_{y-}^{14}} \frac{(-1)^{k_y}}{k_y!}
 \binom{n_{y+}^{14}}{n_{y-}^{14}-k_y}
 \sum_{k'_y = 0}^{n_{y-}^{23}} \frac{(-1)^{k'_y}}{k'_y!}
 \binom{n_{y+}^{23}}{n_{y-}^{23}-k'_y} \Gamma(s_y + k_y + k'_y +
 \unmedio) \nonumber \\
 &   & \mathrm{B}\left( \frac{1}{2}, s_x + s_y + k_x + k_y + k'_x + k'_y +
 \frac{1}{2} \right) \nonumber \\
 &   & \phantom{}_2F_1 \left( s_x + k_x +k'_x + \frac{1}{2},
 \frac{1}{2}; s_x + s_y + k_x + k_y + k'_x + k'_y + 1, 1 -
 \gamma^2 \right).
\end{eqnarray}

\section{Recurrence}
In order to optimize the computing times, let us try to find some
recurrence relations for the kernel of the summation.
\begin{eqnarray}
v_{ij} & = & \mathrm{B} \left( \unmedio, i + j + \unmedio \right)
\phantom{}_2F_1 \left( i + \unmedio, \unmedio; i + j + 1; 1 - \gamma^2
\right) \nonumber \\
 & = & \mathrm{B}_{i+j}F_{i,i+j}.
\end{eqnarray}
The Beta function behaves as follows
\begin{eqnarray}
\mathrm{B}_0 & = & \pi, \label{eq:b1} \\
\mathrm{B}_{k} & = & \frac{k-\unmedio}{k} \mathrm{B}_{k-1}. \label{eq:b2}
\end{eqnarray}
The Gauss hypergeometric function does it in a more complex way
\begin{eqnarray}
a (z - 1) \phantom{}_2F_1(a + 1, b; c;
  z) & = & (a - c) \phantom{}_2F_1(a-1,b;c;z) + (c - 2 a + (a - b) z)
  \phantom{}_2F_1(a, b; c; z) \\
 0 & = & (c - 1) c (z - 1) \phantom{}_2F_1(a, b, c - 1, z) \nonumber
  \\
   &   & + c (c - 1 + (a + b - 2 c + 1) z) \phantom{}_2F_1(a, b, c, z)
  \nonumber \\
   &   & + (a - c) (b - c) z \phantom{}_2F_1(a, b, c + 1, z) \\
\phantom{}_2F_1(a,b;c;z) & = &
\frac{(c-1)\left(2-c-(a+b-2c+3)z\right)}{(a-c+1)(b-c+1)z}
\phantom{}_2F_1(a,b;c-1;z) \nonumber \\ &  & + 
\frac{(c-1)(c-2)(1-z)}{(a-c+1)(b-c+1)z} \phantom{}_2F_1(a,b;c-2;z), \\
\phantom{}_2F_1(a,b;c;z) & = & \frac{c-2a+2+(a-b-1)z}{(a-1)(z-1)}
\phantom{}_2F_1(a-1,b;c;z) \nonumber \\
&   & + \frac{a-c-1}{(a-1)(z-1)} \phantom{}_2F_1(a-2,b;c;z).
\end{eqnarray}
In our case $a=k+\unmedio$, $b=\unmedio$ and $c=l+1$. Therefore if $F_{kl}=\phantom{}_2F_1(k+\unmedio,\unmedio;l+1;z)$:
\begin{eqnarray}
F_{kl} & = & \frac{l (1 - l - (2 + k - 2 l) z)}{(\unmedio -
l)(\unmedio + k - l)z} F_{k,l-1} \nonumber \\
 &   & \frac{l(l -1 )(1 - z)}{(\unmedio-l)(\unmedio+k-l)z}
 F_{k,l-2}, \label{eq:f1} \\
F_{kl} & = & \frac{2-2k+l+(k-1)z}{(k-\unmedio)(z-1)} F_{k-1,l} \nonumber \\
 &   & \frac{k-l-\tresmedio}{(k-\unmedio)(z-1)} F_{k-2,l}, \label{eq:f2}
\end{eqnarray}
where $z=1-\gamma^2$. Now combining expressions (\ref{eq:b2}), (\ref{eq:f1})
and (\ref{eq:f2}), it is possible to find a recursive relation for the
rows and columns of $v_{ij}$. For the columns:
\begin{eqnarray}
v_{ij} & = & \mathrm{B}_{i+j}F_{i,i+j} \nonumber \\
       & = & \frac{i+j-\unmedio}{i+j}\mathrm{B}_{i+j-1} \left(
       \frac{(i + j) \big(1 - i - j - (2  - i - 2 j)
       (1-\gamma^2)\big)}{(j - \unmedio)(i+j-\unmedio)(1-\gamma^2)}
       F_{i,i+j-1} \right. \nonumber \\
       &   & + \left. \frac{(i + j -1) (i + j) \gamma^2}{(j -
       \unmedio)(i+j-\unmedio)(1-\gamma^2)}
       F_{i, i + j - 2} \right) \nonumber \\
       & = & \frac{1 - i - j - (2 - i - 2 j) (1-\gamma^2)}{(j -
       \unmedio)(1-\gamma^2)} \mathrm{B}_{i+j-1}
       F_{i,i+j-1}
       + \frac{(i + j -1) \gamma^2}{(j -
       \unmedio)(1-\gamma^2)} \mathrm{B}_{i+j-1}
       F_{i, i + j - 2} \nonumber \\
       & = & \frac{1 - i - j - (2 - i - 2 j) (1-\gamma^2)}{(j -
       \unmedio)(1-\gamma^2)} v_{i,j-1}
       + \frac{(i + j -1) \gamma^2}{(j -
       \unmedio)(1-\gamma^2)}
       \frac{i+j-\tresmedio}{i+j-1} \mathrm{B}_{i+j-2} F_{i, i + j -
       2} \nonumber \\
       & = & \frac{1- i - j - (2 - i - 2 j) (1-\gamma^2)}{(j -
       \unmedio)(1-\gamma^2)} v_{i,j-1}
       + \frac{\big(i + j - \tresmedio\big) \gamma^2}{(j -
       \unmedio)(1-\gamma^2)} v_{i,j-2},
\end{eqnarray}
And for the rows:
\begin{eqnarray}
v_{ij} & = & \mathrm{B}_{i+j}F_{i,i+j} \nonumber \\
       & = & \mathrm{B}_{i+j} \left( \frac{1+j+(1-i)\gamma^2}{\left(
       \unmedio -i \right)\gamma^2} F_{i-1,i+j} +
       \frac{\left( j + \tresmedio \right)}{\left(i - \unmedio
       \right)\gamma^2} F_{i-2,i+j} \right) \nonumber \\
       & = & \frac{1+j+(1-i)\gamma^2}{\left( \unmedio -i
       \right)\gamma^2} v_{i-1,j+1} +  \frac{j + \tresmedio}{\left(i -
       \unmedio \right)\gamma^2} v_{i-2,j+2}.
\end{eqnarray}
The diagram for the column iteration (with fixed row).
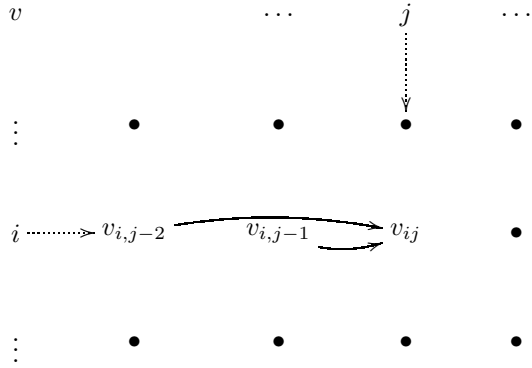
\begin{displaymath}
\xymatrix{
  v & & \ldots & j \ar@{.>}[d] & \ldots\\
  \vdots & \bullet & \bullet & \bullet & \bullet \\
  i \ar@{.>}[r] &
  v_{i,j-2} \ar@/^/[rr] &
  v_{i,j-1} \ar@/_/[r] &
  v_{ij} & \bullet \\
  \vdots & \bullet & \bullet & \bullet & \bullet \\
}
\end{displaymath}
And for row iteration.
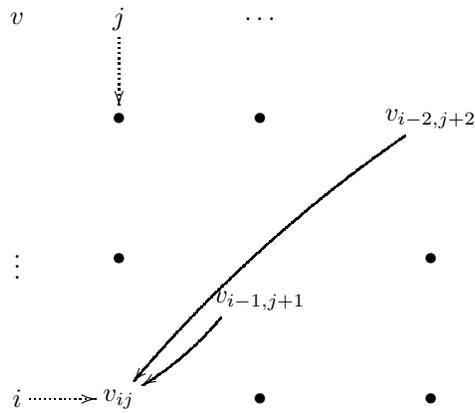
\begin{displaymath}
\xymatrix{
     v   & j \ar@{.>}[d] & \ldots &  \\
         & \bullet & \bullet & v_{i-2,j+2} \ar@/_/[ddll] \\
  \vdots & \bullet & \rule{0pt}{36pt} v_{i-1,j+1} \ar@/^/[dl] & \bullet \\
  i \ar@{.>}[r] & v_{ij} & \bullet & \bullet \\
}
\end{displaymath}
The seed for the iteration is:
\begin{displaymath}
\begin{array}{c|ccc}
v_{ij} & 0 & 1 & 2 \\
\hline
  0    & 2 K(z)  & \frac{2}{z}\big( E(z) + (z-1)K(z) \big)  &   \\
  1    & \frac{2}{z}\left( K(z) - E(z) \right)  &  \frac{2}{z^2}
  \left( 2(z-1)K(z) - (z-2)E(z) \right)  & \\
  2    &  &  & 
\end{array}
\end{displaymath}
\section{Notes}
The Gamma function for half-integer values is defined as:
\begin{equation}
\Gamma \left( n + \unmedio \right) = \frac{(2n-1)!!}{2^n} \sqrt{\pi}.
\end{equation}
\end{document}